\documentclass[twocolumn,showpacs,preprintnumbers,amsmath,amssymb]{revtex4}

\usepackage{graphicx}
\usepackage{dcolumn}
\usepackage{bm}

\newcommand{\disav}[1]{\left\langle #1\right\rangle}
\newcommand{\e}[0]{\text{e}}
\newcommand{\cd}[0]{\text{d}}
\newcommand{\ii}[0]{\text{i}}
\newcommand{\diszeta}[0]{\disav{\zeta}}
\newcommand{\erf}[0]{\text{erf}}

\newcommand{\deph}[0]{h_{\text{p}}}
\newcommand{\downto}[0]{\searrow}

\begin{document}


\title{On mean field theory for ac-driven elastic interfaces exposed to disorder}

\author{Friedmar~Sch\"utze}%
 \email{schuetze@thp.uni-koeln.de}
\affiliation{%
Institut f\"ur Theoretische Physik, Universit\"at zu K\"oln,
Z\"ulpicher Stra\ss e 77, 50937 K\"oln, Germany
}%

\date{\today}

\begin{abstract}
The analytic description of ac-driven elastic interfaces in random potentials
is desirable because the problem is experimentally relevant. 
This work emphasises on the mean field approximation
for the problem at zero temperature.
We prove that perturbation theory is regular in all orders by giving an 
inductive scheme how to find groups of ill-behaved graphs that mutually cancel,
leaving a regular expression. In the parameter regimes for which perturbation theory
is applicable it agrees with numerical results. 
Further, we determine the dependence of
the 
Fourier coefficients of the mean velocity on
the parameters of the model.
\end{abstract}

\maketitle

\section{\label{sec:intro}%
Introduction
}
The theory for elastic interfaces in a disordered
environment, driven by an external dc-force at temperature $T=0$ is
widely understood, and also in the finite temperature 
case some progress has been achieved
\cite{NB:AdvPh04,Middleton:PRB92,Lemetal:PRL98,CGD:PRB00,Metetal:PRL07,BKG:EPL08}. 
At $T=0$ the dc driven interface exhibits an interesting critical point, corresponding to
the depinning transition. For small constant driving forces $h$, the interface
adjusts its configuration to balance the driving force and the disorder, but
remains pinned and does not move on large time scales. If $h$ reaches a critical
threshold $\deph$, the interface starts to slide with a mean velocity $v$ that behaves
as $v\sim(h-\deph)^\beta$ for $h\downto\deph$.
The critical properties of this non-equilibrium transition have
been worked out by the use of functional renormalisation group  methods 
\cite{NSTL:JP2F,NarayanFisher:PRB92,Ertas:PRE94,NSTL:APL,CDW:PRL01,CDW:PRB02,CDW:PRE04}.

Beyond constant driving forces, 
experimental achievements on the problem of ac-driven
elastic domain walls in ferroic systems 
\cite{Kleemann:PRL07,Kleemann:ARMR07,Jezewski:PRB08}
emphasise the importance of a theoretical understanding.
Despite the experimental progress, and in contrast to the
success in understanding the dc case, yet
there is little advance in the theoretical description of the
problem, even for $T=0$.
The exact solution of the equation of motion is deemed impossible
due to the complicated non-linear feedback of the interface's
configuration in the disorder force term.
To make matters worse, attempts to access the problem for an ac-driving force
perturbatively by an expansion in the disorder strength 
bring along severe problems \cite{FSpert09}.

This underlines the importance of the mean-field approach,
which is the central subject of this article.
We investigate the behaviour of ac-driven domain walls in a disordered
environment in the mean-field approximation and prove, 
that the perturbative corrections remain bounded in all orders.
Further, we indicate, that for large enough driving field amplitudes,
sufficiently strong elastic coupling and high frequencies, 
the perturbative
results agree very well with the numerics for the full mean field equation
of motion. 
The quantitative statements that rely on a special choice for the disorder 
correlator are worked out for
elastic manifolds, like for example interfaces 
between two immiscible fluids or domain walls in ferroic systems, 
exposed to random field disorder.
Our proof of
the regularity of perturbation theory should also extend 
to similar models that describe other interesting physical systems 
with disorder
\cite{Kardar:PhRep98,Fisher:PhRep98},
for example charge-density waves \cite{GruenerCDW:RMP88} 
and flux lines in type-II superconductors 
\cite{Blatter:RMP94,GiaDou:PRL94,NatSchei:AdvPh00}.

Mean field calculations have been performed for similar problems
before.
D.S. Fisher \cite{Fisher:PRL83,Fisher:PRB85} 
calculated dynamic properties of sliding charge-density waves
in a mean-field model with dc-driving and argued in favour of a depinning transition
in the strong pinning case. Using smooth bounded disorder, he calculated
the threshold field for depinning as well as critical exponents related
to the depinning transition. Furthermore, he considered
the response in case of an ac-field applied in addition to the dc-driving.
The perturbation expansion for dc-driven interfaces has been 
investigated by Koplik and Levine \cite{KL:PRB85}, who also emphasised on the
mean-field problem. 
Later, Leschorn \cite{Leschorn:JPA92} calculated the depinning force
and the critical properties of the depinning transition for a 
three-state random field model. Narayan and 
Fisher \cite{NarayanFisher:PRB92} investigated the critical behaviour of
charge density waves in the sliding regime and worked out the threshold
field for scalloped disorder potentials. Lyuksyutov considered
dynamical friction and instability phenomena for the interface motion
\cite{Lyuksyutov:JPC95}.

The rest of this article is organised in the following way:
In the next section, we are going to deduce the mean-field equation of
motion from the original model taken to describe the problem of
disordered elastic domain walls. In section \ref{sec:pert}, we establish the
diagrammatic perturbation expansion for the mean-field theory and
show its regularity. The bulk part of the inductive proof (the induction step)
is outsourced to the appendix. After a brief review of the
problem for a constant driving force in section \ref{sec:adiabatic}, we
focus our attention to the ac-driving case in section \ref{sec:ac}.
There, we start with some qualitative discussion of the numerical
solution and then go on to
present our attempts to extract information from the first non-vanishing
perturbative terms. Due to the complicated non-linear structure of the
expressions involved, numerical methods had to be employed as well. However,
the perturbative approach helps a lot to improve numerical results.
This makes it possible to work out the decay law of the Fourier coefficients with
their order, in dependence of the strength of the driving field.

\section{\label{sec:model}%
The model%
}


To model $D$-dimensional elastic manifolds in a $D+1$-dimensional
disordered system, we employ an equation of
motion that has been introduced in a number of earlier works
\cite{Feigelman:JETP83,Bruinsma:PRL84,KL:PRB85}
\begin{equation}
\label{eq:weom}
\partial_tz(x,t)=\Gamma\nabla_x^2z+h\cdot f(t)+
u\cdot g(x,z).
\end{equation}
The equation does not involve a thermal noise term and therefore describes
the (classical) system at $T=0$.
The interface profile is described by the single-valued function 
$z(x,t)$, so we do not allow for overhangs. Here, $x$ is the 
$D$-dimensional set of coordinates which parameterise the interface 
manifold itself. $\Gamma$ denotes the elastic
stiffness constant of the interface, $h$ measures the strength of the
external driving force and $f(t)$ denotes its time-dependence, taken to
be of order unity. We did not specify $f(t)$ yet to allow for general 
considerations. The disorder is modeled by $g(x,z)$, its strength is
measured by $u$. We assume quenched Gau\ss ian disorder, characterised by
its first two cumulants:
\begin{eqnarray}
\label{eq:wdisorder}
\disav{g(x,z)}&=&0\nonumber,\\
\disav{g(x,z)g(x',z')}&=&\delta^D(x-x')\Delta(z-z').
\end{eqnarray}
Here, $\disav{\ldots}$ denotes the average over disorder.
The function $\Delta(z)$ will be specified further down.
As the short-ranged elastic term suggests, long-range interactions, of
dipolar type for example, are not covered.

The corresponding mean field equation (cf. e.g. \cite{Fisher:PRB85}) 
is obtained via the replacement of the elastic term
by a uniform long-range coupling. To do this,
we have to formulate the model (\ref{eq:weom}) on a lattice in
$x$-direction, i.e. the coordinates that parameterise the interface itself 
are discretised.
The lattice Laplacian reads
\begin{align}
\nabla_x^2z(x_i)&=
\sum\limits_{d=1}^D{z(x_i+ae_d)+z(x_i-ae_d)-2z(x_i)\over a^2}
\nonumber\\
&=\sum\limits_{d=1}^D\sum\limits_{j_d=1}^NJ_{ij_d}\big[z(x_{j_d})-z(x_i)\big],
\nonumber\\
J_{ij_d}&={1\over a^2}\big[\delta_{j_d+1,i}+\delta_{j_d-1,i}\big]\nonumber,
\end{align}
where $a$ denotes the lattice constant.
To get the mean field theory, $J_{ij}$ has to be replaced by a uniform coupling but
such that the sum over all couplings $\sum_jJ_{ij}$ remains the same. Hence, we choose
\begin{align}
J_{ij}^{\rm MF}={1\over a^2N}.
\end{align}
Now, the disorder has to be discretised as well, which is achieved if we
replace the delta function in the correlator (\ref{eq:wdisorder}) by
$\delta^D(x_i-x_j)\to \delta_{ij}a^{-D/2}$ (cf. \cite{Bruinsma:PRL84}).
The resulting equation of motion should be independent of $x$, just the 
lattice constant $a$ and the
dimension enter because the disorder scales with a factor $a^{-D/2}$. 
Finally, for the mean-field equation of motion, we obtain
\begin{align}
\label{eq:mfgeom}
\partial_tz
&=c\cdot\left[\disav{z}-z\right]+h\cdot f(t)+
\eta\cdot g(z),
\end{align}
where $c=\Gamma/a^2$ and $\eta=u/a^{D/2}$.
We assume quenched Gau\ss ian disorder with
\begin{eqnarray}
\label{eq:disorder}
\disav{g(z)}&=&0\\
\disav{g(z)g(z')}&=&\Delta(z-z').
\end{eqnarray}
The function $\Delta(z-z')$ shall be
smooth, symmetric and should decay exponentially
on a length scale $\ell$. Moreover,
we require $\Delta(0)=1$, as the disorder strength shall be measured by
$\eta$. For the sake of concreteness, we shall choose
\begin{equation}
\label{eq:discorr}
\Delta(z-z')=\exp\left[-\left({z-z'\over\ell}\right)^2\right]
\end{equation}
whenever we need an explicit expression for calculations. This
disorder correlator correctly describes the situation for an elastic manifold
in random field disorder \cite{NSTL:APL}, sufficiently far away from the
critical depinning transition point.

The physical picture of the
mean field equation of motion is a system of distinct particles,
moving in certain realisations of the disorder. All of them are
harmonically coupled to
their common mean, i.e. the elastic coupling between neighbouring
wall segments $\Gamma\nabla_x^2z$ is now replaced by a uniform coupling
$c\cdot\left[\disav{z}-z\right]$ to 
the disorder averaged position $\disav{z}$, which in turn is determined self-consistently 
by the single realisations.

%
Apart from the correlation length $\ell$ of the disorder, there is another
important length scale in the system. In the absence of any
driving force (i.e. $h=0$), we can easily determine the
mean deviation of the coordinate
$z$ of a special realisation from the disorder averaged position 
$\disav{z}$. For $h=0$ we expect $\dot z=0$, at least in the steady state
and (\ref{eq:mfgeom}) straightforwardly leads to 
$$
\disav{(\disav{z}-z)^2}\simeq {\eta^2\over c^2}.
$$
So, $\eta/c$ measures the modulus of the average distance from the common mean.
We will see below, that $\eta/c$ is an 
upper bound in the case, that the system moves under the influence of a
non-zero driving $h\ne 0$. 

A word on notation: The disorder averaged velocity $v=\disav{\dot z}$ will
be denoted by the symbol $v$.

\section{\label{sec:pert}%
Perturbation theory%
}

\subsection{\label{sec:pert:diag}%
Diagrammatic expansion%
}

The differential equation of motion (\ref{eq:mfgeom}) is non-linear,
and due to the influence of the solution on the disorder
it is impossible to solve it exactly. An ansatz is, to attempt an expansion
in the disorder strength $\eta$. Therefore, we decompose $z=Z+\zeta$, where 
$Z=hF(t)$ (with $\partial_tF(t)=f(t)$) is
the solution of the non-disordered problem ($\eta=0$) around which we expand, and 
$$
\zeta=\sum\limits_{k=1}^\infty\zeta_k\eta^k\>,\quad
\diszeta=\sum\limits_{k=1}^\infty\diszeta_k\eta^k.
$$
is the perturbative correction.
Still, we have the equations for $\zeta_k$ depending on $\diszeta_k$,
which is also unknown. This eventually leads us to a set of two coupled
equations
\begin{align}
\label{eq:mfpertsys1}
(\partial_t+c)\zeta&=c\diszeta+\eta\cdot g(Z+\zeta)\\
\label{eq:mfpertsys2}
\partial_t\diszeta&=\eta\cdot\disav{g(Z+\zeta)},
\end{align}
that we can solve iteratively for every order of the perturbation 
series, if we expand
\begin{align}
\label{eq:mfdisexpansion}
g(Z+\zeta)=\sum\limits_{n=0}^\infty{g^{(n)}(Z)\over n!}\zeta^n.
\end{align}
If one is interested to keep small orders, this expansion of the disorder 
can only work if $\zeta\ll\ell$, because $\ell$ is the typical scale on which $g(z)$
changes. 
We will come back to that point, when discussing the
special cases for $f(t)$ in sections \ref{sec:adiabatic} and \ref{sec:ac}. 
For the moment, we just do it.

The propagator corresponding to the left hand side of Eq. (\ref{eq:mfpertsys1}) reads
$$
G(t)=\Theta(t)\cdot\e^{-ct}.
$$
Using this propagator, we can formally write down the solution and express
it order by order in a power series in $\eta$.
Up to the second order, the solutions are
\begin{align}
\diszeta_1(t)&=0,\\
\zeta_1(t)&=\int\limits_0^t\cd t_1\>\e^{-c(t-t_1)}g(Z(t_1)),\\
\label{eq:diszeta2}
\diszeta_2(t)&=\int\limits_0^t\cd t_1\int\limits_0^{t_1}\cd t_2
\>\e^{-c(t_1-t_2)}\Delta'[Z(t_1)-Z(t_2)],\\
\label{eq:zeta2}
\zeta_2(t)&=\int\limits_0^t\cd t_1\e^{-c(t-t_1)}\left[c\diszeta_2(t_1)+g'(Z(t_1))
\cdot\zeta_1(t_1)\right].
\end{align}%
Since we assume Gau\ss ian disorder, the disorder averaged corrections
$\diszeta_n$ vanish for odd $n$.
Due to the nested structure, a diagrammatic representation of the perturbation 
expansion seems most suited. For the interesting quantities $\diszeta_k$, 
the first two non-vanishing orders are given by: 
\begin{widetext}
\begin{align}
\label{eq:diagexpand}
\diszeta_2=&\quad
\parbox{17mm}{\includegraphics{zdiag.1}}
\nonumber\\&\phantom{=}\nonumber\\
\diszeta_4=&\quad 
3\cdot\parbox{17mm}{\includegraphics{zdiag.2}}\quad+\quad
\parbox{25mm}{\includegraphics{zdiag.3}}\quad+\quad
2\cdot\parbox{25mm}{\includegraphics{zdiag.4}}\quad+
\nonumber\\&\phantom{=}\\&\quad
2\cdot\parbox{25mm}{\includegraphics{zdiag.5}}\quad+\quad
2\cdot\parbox{25mm}{\includegraphics{zdiag.6}}\quad+\quad
2\cdot\parbox{25mm}{\includegraphics{zdiag.7}}\quad+
\nonumber\\&\phantom{=}\nonumber\\&\quad
\parbox{33mm}{\includegraphics{zdiag.8}}\quad+\quad
\parbox{33mm}{\includegraphics{zdiag.9}}\quad+\quad
\parbox{33mm}{\includegraphics{zdiag.10}}
\nonumber
\end{align}
\end{widetext}
The diagrammatic rules are fairly simple: 
we draw all rooted trees with $k$
vertices, and add a stem. Each vertex
corresponds to a factor $g^{(m)}(Z(t))/m!$, where $m$ counts the number of
outgoing lines (away from the root). 
The line between two vertices represents a propagator $G(t)$.
Then Wick's theorem is applied to carry out the disorder average. Each two vertices, that
are grouped together for the average, will be connected by a dashed line.
Finally, we replace every straight line which, upon removing it, makes the whole
graph falling apart into two subgraphs, by a curly line, corresponding to 
the propagator of (\ref{eq:mfpertsys2}), which is just a Heaviside
function $\Theta(t)$. In the framework of equilibrium quantum statistical physics, 
those graphs that involve an internal curly line are called one-particle reducible (1PR).
In our classical problem loops only occur due to the dashed lines originating from 
the Gau\ss ian average.

\subsection{\label{sec:pert:consistency}%
Consistency of the perturbative series%
}
The perturbation expansion leaves some questions, that have to be addressed.
It is not immediately obvious, that taking the disorder average of (\ref{eq:zeta2})
gives the result in (\ref{eq:diszeta2}), i.e. $\diszeta_2(t)=\disav{\zeta_2(t)}$.
However, a short calculation, using integration by parts reveals this relation to hold.

Another, much deeper problem is related to the diagrams involving a curly line in 
their interior.
Due to the curly line, they grow linearly in time. In the following, we call diagrams
non-regular, if they correspond to terms which grow unboundedly in time. Koplik and
Levine \cite{KL:PRB85} explicitly checked for a time independent driving up to sixth
order, that the problematic terms of the diagrams mutually cancel. 
We give a very general version, that holds for any $f(t)$ and covers all perturbative
orders. To illustrate, how this works, we present the calculation for the
fourth order here. The somewhat technical induction step, which extends our argument
to all orders is given in appendix \ref{app:reg}.
For simplicity, we work with the velocity diagrams, that are obtained by just
removing the curly line from the root.
\begin{widetext}
\begin{align}
2\cdot\>\parbox{16mm}{\includegraphics{vdiag.1}}=&\int\limits_0^t\cd t_1\>\e^{-c(t-t_1)}
\Delta''[Z(t)-Z(t_1)]\int\limits_0^t\cd t_2\int\limits_0^{t_2}\cd t_3\>
\e^{-c(t_2-t_3)}\Delta'[Z(t_2)-Z(t_3)]\nonumber\\
\parbox{24mm}{\includegraphics{vdiag.2}}=&\int\limits_0^t\cd t_1\>\e^{-c(t-t_1)}
(-\Delta''[Z(t)-Z(t_1)])\int\limits_0^{t_1}\cd t_2\int\limits_0^{t_2}\cd t_3\>
\e^{-c(t_2-t_3)}\Delta'[Z(t_2)-Z(t_3)]\nonumber\\
=&-2\cdot\>\parbox{16mm}{\includegraphics{vdiag.1}}+S\nonumber\\
S=&\int\limits_0^t\cd t_1\int\limits_0^{t_1}\cd t_2\>\e^{-c(t-t_2)}
\Delta''[Z(t)-Z(t_2)]\int\limits_0^{t_1}\cd t_3\>
\e^{-c(t_1-t_3)}\Delta'[Z(t_1)-Z(t_3)]\nonumber
\end{align}
\end{widetext}
The modification of the second diagram to express it as the sum of the first and $S$
is merely integration by parts.
The term $S$ now corresponds to the sum of the two diagrams. It is not 
straightforward to see, how $S$ behaves generally, but it is easy to see, that it
remains bounded for large times. Every time integral carries an exponential
damping term.
Basically, we have thereby established, that at least up to the fourth order, the
perturbation series exists and is well-behaved in the sense, that there are
no terms that lead to an overall unbounded growth in time.

Our analysis how the cancellations among non-regular diagrams (i.e. those
that involve an internal curly line) generalise to higher orders is presented in
appendix \ref{app:reg}.

\section{\label{sec:adiabatic}%
Time-independent driving force%
}
Let us consider the mean field equation (\ref{eq:mfgeom}) for the special
choice $f(t)=1$. Particular cases of this problem have already been 
addressed
\cite{Fisher:PRL83,Fisher:PRB85,KL:PRB85,Leschorn:JPA92,NarayanFisher:PRB92}. 
Assuming that the disorder correlator is cusped, e.g.
\begin{equation}
\label{eq:cdiscorr}
\Delta_{\rm c}(z-z')=\exp\left[-{|z-z'|\over\ell}\right],
\end{equation}
it is expected, that the system, described by (\ref{eq:mfgeom})
shows a depinning transistion. 
The special feature of a cusped disorder potential is, that the
resulting disorder force $g(z)$ exhibits jumps. At such jumps,
the system is pinned and a finite threshold force is needed to 
move it in a certain direction.
The critical depinning force
has been determined in \cite{NarayanFisher:PRB92} for the
parabolic scalloped potential and for
a three-state random field model in \cite{Leschorn:JPA92}. In both works,
the critical exponent was found to be $\beta=1$.

For a disorder correlator given by (\ref{eq:discorr}), numerical investigations
suggest, that pinning is exponentially suppressed for large $c$. More precisely,
the depinning field $\deph$ obeys
$\deph\propto\exp[-C\cdot c\ell/\eta]$ with some numerical prefactor $C$.
This is reasonable, since $\eta/c$ is a measure for the typical deviation 
of each single realisation from the mean, so for $\eta/c\ll\ell$, the system
is prevented from adopting to the minima of the disorder potential
by the elastic force. Thus, pinning should be diminished. On the
other hand, for $\eta/c>\ell$, the equilibrated system for $h=0$
does place itself at the local disorder minima 
(for all realisations), since on a scale $\ell$ it is expected to find
a local minimum. So the existence of a finite perceptable 
threshold force $\deph$ is possible for small enough $c$.
Fisher \cite{Fisher:PRB85} analytically found a threshold field for bounded
disorder (strong enough) in the case of charge density waves. The critical
exponent has been found to be $\beta=3/2$ for the smooth disorder potential, 
in contrast with $\beta=1$ for the cusped case. Thus, as pointed out in
\cite{NarayanFisher:PRB92}, the exponents for the depinning
transition in the mean-field case are non-universal.

Well above the depinning threshold, i.e. for
$h\gg\deph$, perturbation theory should give a good
estimate of the mean velocity. 
However, it is not clear, why our perturbative approach via a
systematic expansion in the disorder (essentially equal to the
calculations in \cite{KL:PRB85}) can work, 
since truncating the Taylor expansion (\ref{eq:mfdisexpansion})
at a finite order is only a good estimate, if $\zeta\ll\ell$.
But, the true velocity $v$ is certainly different from $h$, 
thence $\zeta=(v-h)t$ grows linearly in time. 
In appendix \ref{app:resum} we illustrate, how a resummation of the
perturbation scheme leads to a perturbative 
programme that works. 
To take only small orders into account
a necessary condition is that $\eta/c\ll\ell$.
Choosing (\ref{eq:discorr}) for the disorder correlator,
the first non-vanishing order reads
\begin{align}
\label{eq:pert4o}
{v\over h}=&1-{\eta^2\over vh}\left[1-\phi\left({c\ell\over 2v}
\right)\right]
+{\cal O}\left({\eta^4\over hv^3}\right),
\end{align}
where we have introduced the function
\begin{equation}
\label{eq:phifkt}
\phi(x)=\sqrt{\pi}\cdot x\>\exp\left({x^2}\right)\>\left[1-\erf(x)\right]
\end{equation}%
for convenience. Its asymptotic expansion reads
\begin{align*}
\phi(x)&=\left\lbrace
\begin{matrix}
\sqrt{\pi}x-2x^2+{\cal O}(x^3)&x\ll 1\\
1-{1\over 2x^2}+{\cal O}(x^{-4})&x\gg 1
\end{matrix}\right.\,.
\end{align*}
For $\ell\to\infty$, in (\ref{eq:pert4o}) all perturbative corrections vanish,
hence we get $v=h$. This was expected, since, if the disorder force is correlated over 
an infinite range, it is essentially constant. Taking the average over all possible
values of the disorder (positive and negative) gives zero, hence there is no
disorder effect any more. For finite $\ell$ the velocity is reduced, 
the smaller $c\ell$, the more. For large $h\gg\deph$, 
the difference $h-v$ is expected to be small,
hence it does not cause much harm if one replaces $v$ by $h$ on the right hand side
of (\ref{eq:pert4o}), rearriving at the explicit expansion in the disorder.

The results from above mainly agree with those in \cite{KL:PRB85}.
The major difference is, that Koplik and Levine
have been working with $\Delta_{\rm c}/(2\ell)$ instead of (\ref{eq:discorr})
and therefore
get an expansion in $\eta^2/(\ell h^2)$, whereas we have a power series
in $\eta^2/h^2$.

\section{\label{sec:ac}%
Considerations for ac driving forces%
}

\subsection{\label{sec:ac:numerics}%
Qualitative behaviour and numerical results%
}
To get an idea about how the system, corresponding to the equation
of motion with an ac-driving (cf. (\ref{eq:mfgeom}))
\begin{align}
\label{eq:acmfeom}
\partial_tz=c\cdot[\disav{z}-z]+h\cdot\cos\omega t+\eta\cdot g(z)
\end{align}
behaves, we implemented a numerical approach. The disorder is modelled
by concatenated straight lines, the values of the junction points are 
chosen randomly from a bounded interval. The correlator has been checked to
be perfectly in agreement with (\ref{eq:discorr}).

\begin{figure}
\includegraphics[width=0.9\columnwidth]{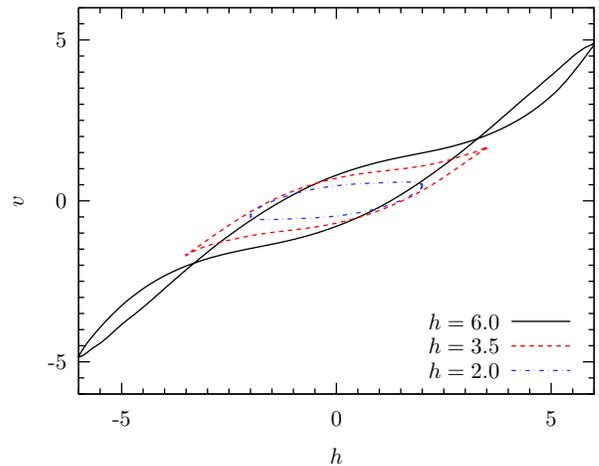}
\caption{\label{fig:qbild} Numerical solution of (\ref{eq:acmfeom})
for different driving field strengths and $c=1.0$, $\eta=2.5$.
For the simulation, $t$ and $z$ are measured in units such that
$\omega=\ell=1$.}
\end{figure}%
Before discussing the numerical trajectories, we note a first property
of the equation of motion (\ref{eq:acmfeom}). It contains a symmetry of the
(disorder averaged) system, namely that all disorder averaged
quantities are invariant under the 
transformation $h\to -h$ 
and $z\to -z$, which implies $v\to -v$. We have hereby fixed the initial
condition to be $z(0)=0$ for all realisations. If one chooses another
initial condition, its sign has to be inverted as well, of course.
In the steady state, i.e. for $t\gg c^{-1}$ (as $c^{-1}$ is the 
time-scale on which transience effects are diminished, see below), the
trajectory must therefore obey the symmetry $h\to -h$, $v\to-v$.
This symmetry is obviously reflected in the numerical solutions 
(see fig. \ref{fig:qbild}).

An interesting consequence of this symmetry is, 
that the even Fourier coefficients of
the solution $v(t)$ (which is periodic with period $2\pi/\omega$)
vanish. Once the steady state is reached, the symmetry requires
$v(t)=-v(t+\pi/\omega)$. For the even Fourier modes this means
\begin{align*}
c_{2N}&=\int\limits_0^{2\pi\over\omega}\cd t\>v(t)\e^{\ii2N\omega t}\\
&=\int\limits_0^{\pi\over\omega}\cd t\>v(t)\e^{\ii2N\omega t}+
\int\limits_0^{\pi\over\omega}\cd t\>v\left(t+{\pi\over\omega}\right)
\e^{\ii2N\omega t}
=0.
\end{align*}

The typical picture of a $v$-$h$-plot is that of a single hysteresis
for $h\ll\eta$ and  a double hysteresis for $h\gg\eta$. In an intermediate
range, we find a single hysteresis with a cusped endpoint. The qualitative
shape of the solution trajectories agrees with numerical 
results \cite{GNP:PRL03,Glatz:phd}, that
have been obtained as solutions for (\ref{eq:weom}) in the case of finite
interfaces with periodic boundary conditions.
\begin{figure}
\includegraphics[width=\columnwidth]{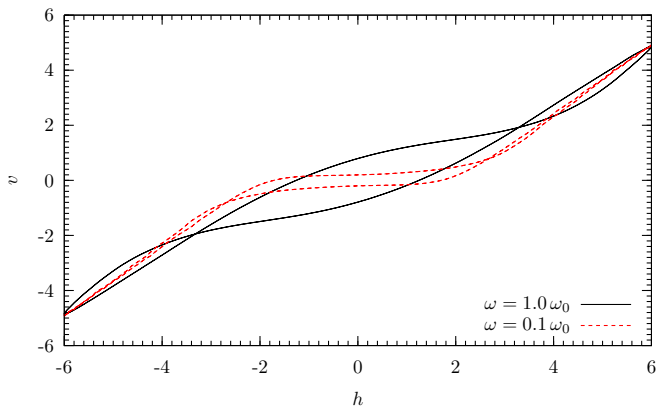}
\caption{\label{fig:wto0} Numerical solution of equation (\ref{eq:acmfeom})
for $h=6.0$, $c=1.0$ and $\eta=2.5$ for different frequencies, 
$t$ and $z$ being measured in units such that
$\omega_0=\ell=1$.}
\end{figure}%
Moreover, as the frequency is sent to zero $\omega\to 0$, the hysteretic
trajectory approaches the depinning curve for an adiabatic
change of the driving field. This is shown in fig. \ref{fig:wto0}.

In the following, we want to give a qualitative discussion of the 
hystereses in the case of small elasticity $c$.

\emph{Weak fields $h\ll\eta$:}\quad
In the case of weak driving fields, the typical system in a certain
disorder configuration remains in a potential well of the disorder.
The elastic force may slightly shift the centre around which $z(t)$
oscillates, but this is not
very important, since we can instead think of an effective potential.
To understand the hysteretic behaviour, it is instructive to think of
the force field $g(z)$ instead of the potential. Starting at
$h(t)=0$ for large enough $t$ (i.e. in the steady state) 
we expect a certain realisation to be located at the zero
point $g(z_0)=0$ of a falling edge, since this corresponds to a stable
configuration. As the field grows, the system starts to move in the
direction of growing $z$, where the disorder force competes with
the driving. Because in the vicinity of the potential minimum,
the disorder force $g(z)$ behaves approximately
linear in $z$, the acceleration is approximately zero and
the velocity almost constant. This changes when the driving is about to 
reaching its maximum. The slower the growth, the smaller the velocity.
At the maximum, the velocity equals zero, as the driving and the restitutional 
disorder force compensate. For decreasing $h(t)$, the restitution force
wins and pushes the system back in the direction of the potential minimum. 
Hence, the velocity $v$
turns negative short-time after the field has reached its maximum and is still positive.
Once the stable position $z_0$ is reached again, 
the same starts in the negative direction.

Certainly, the restitutional disorder force need not continuously grow with 
$z$, but may exhibit bumps or similar noisy structure, but those details average 
out when taking the mean over all disorder configurations.

\begin{figure}
\includegraphics[width=\columnwidth]{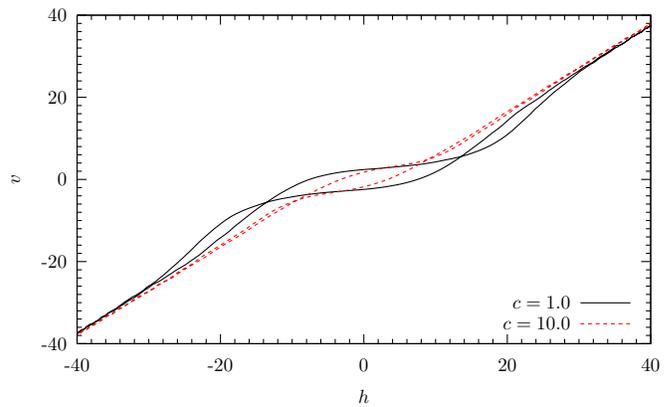}
\caption{\label{fig:ceinfl} Numerical solution of equation (\ref{eq:acmfeom})
for different elastic constants and $h=40.0$, $\eta=10.0$.
The units of $t$ and $z$ are chosen such that
$\omega=\ell=1$.}
\end{figure}

\emph{Strong fields $h\gg\eta$:}\quad
In the case of strong driving amplitudes, we encounter the situation of a
double hysteresis. Again, starting at $h(t)=0$ for $t\gg c^{-1}$,
we assume the system to be
located at the zero $g(z_0)=0$ of a falling edge of the (effective) 
disorder force field. As $h(t)$ grows, we first have the same situation as in
the case of weak driving: The disorder acts restitutionally and thus keeps the 
velocity small and leads to a small slope $\cd v/\cd h$. Once the field
is of the order $\eta$, 
the typical maximum of a disorder force, the system is no longer
locked into a potential well, but a cross-over to sliding behaviour sets in. 
On further increasing $h$, the
system finally arrives at a slope $s=\cd v/\cd h$, which depends mainly
on $\eta$ and $c$. After the field reaches
its maximum, the velocity decreases with the field, the slope being 
$\cd v/\cd h\approx s$, also if this slope has been different just before the 
field amplitude has been reached. This slope approximately remains, until
the field is weaker than the typical disorder force, when the system is
again trapped in a potential well. Since on rising edges of the disorder force,
driving and disorder point in the same direction, the system will rarely
sit there (it moves away very fast). The velocity becomes negative before
$h=0$, since the system slides down the falling edge ($\cd g/\cd z<0$) 
of the disorder force.
At $h=0$ everything starts again in the negative direction.
An example for fairly large
field amplitudes is shown in fig. \ref{fig:ceinfl}.

So far, our discussion has emphasised on small $c$ by absorbing its effect into
an effective disorder picture. The effect of larger $c$ is to couple the 
configuration $z(t)$ of every realisation strongly to the mean $\disav{z(t)}$.
This wipes out the effect of disorder in the time regime when $h(t)$ takes
on small values. As we have discussed above, in those time intervals the
possibility to explore the shape of the individual disorder landscape plays 
an important r\^ole. Thus for larger $c$ the double hysteresis winds around a 
straight line, connecting the extremal velocities. This can be seen in
fig. \ref{fig:ceinfl}.

\subsection{\label{sec:ac:valpert}%
Validity of perturbation theory%
}
For an oscillating driving force, the question is still open, whether one may assume
$\zeta$ to be small compared to $\ell$. 
If $c$ is large, any particle moving in a particular realisation of a
disorder potential is strongly bound to the disorder averaged position.
This prevents it from exploring the own disorder environment and thus
large $c$ effectively scale down $\eta$. All realisations stay close to the
disorder averaged position, the mean deviation being $\eta/c$. 
A problem now occurs, if the disorder averaged
position deviates strongly from the $\eta=0$ solution. For $h\gg\eta$ this can
only happen during those periods, where $h\>\cos\omega t$ takes on small values.
The time, that has to elapse, until every system has adopted to its
own disorder realisation, and hence the time until the system can be pinned,
is $c^{-1}$ (see below). For perturbation theory to work, 
this time must be large compared to the length of the period
during which $h\le\eta$, which we roughly estimate as $\eta/(\omega h)$. This
gives us a second condition for the applicability of perturbation theory:
$h/\eta\gg c/\omega$.

In summary, the conditions for perturbation theory to hold are the
following. The driving force amplitude $h$ has to be large compared to 
$\eta$, $h/\eta\gg\max\{c/\omega,1\}$ to make the series 
expansion work and
to guarantee that the disorder averaged solution stays close to the
$\eta=0$ trajectory (around which we expand). 
Moreover, $c$ must be large ($c\gg\eta/\ell$)
to ensure proximity of each realisation to the disorder average.

\subsection{\label{sec:ac:width}%
Simple perturbative estimates%
}
Before embarking on conclusions that can be drawn from the perturbative
solution of the equation of motion (\ref{eq:mfgeom}), we shall determine
the typical
deviation of the position of a single realisation $z(t)$ 
from the mean $\disav{z(t)}$. This has the following bound
\begin{align}
\disav{(\disav{z}-z)^2}&=\eta^2\disav{\zeta_1^2}+{\cal O}(\eta^4)\nonumber\\
&=\eta^2\int\limits_0^t\cd t_1\cd t_2\>\e^{-c(2t-t_1-t_2)}\Delta[Z(t_1)-Z(t_2)]\nonumber\\
&\le\int\limits_0^t\cd t_1\cd t_2\>\e^{-c(2t-t_1-t_2)}={\eta^2\over c^2}
\left(1-\e^{-ct}\right)^2.
\end{align}
The estimate simply replaces the disorder correlator $\Delta$ by its maximum
and therefore gives an upper bound. Strictly, it is only true to order
${\cal O}(\eta^2)$. This confirmes our claim from section \ref{sec:model}, that
an estimate for the mean deviation from the averaged solution is given 
by $\eta/c$.

\subsection{\label{sec:ac:harmon}%
Perturbative harmonic expansion%
}
\begin{figure}[t]
\includegraphics[width=\columnwidth]{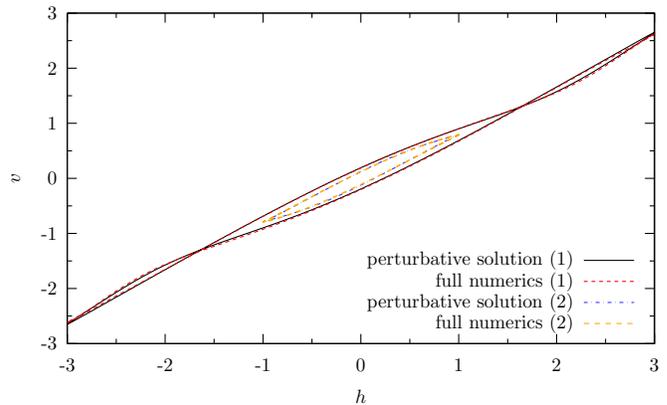}
\caption{\label{fig:stvollvgl} Comparison of the numerical
solution of equation (\ref{eq:acmfeom}) with the result obtained
from the first non-vanishing perturbative order for
(1) $h=3.0$, $c=3.0$, $\eta=1.5$ and 
(2) $h=1.0$, $c=1.0$, $\eta=0.6$.
The units of $t$ and $z$ are chosen such that
$\omega=\ell=1$.}
\end{figure}%
As has been discussed in section \ref{sec:ac:valpert}, for $h/\eta\gg\max\{c/\omega,1\}$ 
and $\eta/(c\ell)\ll 1$, perturbation theory should do pretty well. A direct
comparison, shown in fig. \ref{fig:stvollvgl}, confirms an excellent 
agreement. So, at least for weak disorder, when perturbation theory is 
valid, one can hope to extract some information from the lowest
order.
For an ac driving force, even this lowest perturbative order for the
velocity is a very complicated expression. The diagrammatic prescription
yields up to the order ${\cal O}(\eta^2)$
\begin{equation}
\label{eq:v1stop}
v(t)=h\cos\omega t+\eta^2\int\limits_0^t\cd t'\>
\e^{-c(t-t')}\Delta'[Z(t)-Z(t')].
\end{equation}
Remember, that $Z(t)=(h/\omega)\sin\omega t$ is the solution for the problem without
disorder, around
which we expand.
It seems reasonable to aim a harmonic expansion of the mean velocity $v$.
The ansatz therefore is
\begin{align}
v(t)=\sum\limits_{N=1}^\infty\big[a_N\cos N\omega t+b_N\sin N\omega t\big],
\end{align}
for $N$ odd. Recall, that, for reasons of the $h\to-h$ and $v\to-v$ symmetry
of the trajectory, which has been discussed in the previous section, the
Fourier coefficients for even $N$ vanish.

Starting from the first order result for $v$ (\ref{eq:v1stop}),
we express the disorder correlator by its Fourier transform
\begin{align}
\Delta'[Z(t)-Z(t')]=\int\limits{\cd q\over 2\pi}(\ii q)\Delta(q)
\e^{\ii q{h\over\omega}[\sin\omega t-\sin\omega t']}
\end{align}
and expand the exponential term in a double Fourier series
in $t$ and $t'$, respectively:
\begin{eqnarray*}
\e^{\ii a\sin\omega t}&=&\sum\limits_{n=-\infty}^\infty
J_n(a)\e^{\ii n\omega t}\\
\int\limits_0^t\cd t'\>\e^{-c(t-t')-\ii a\sin\omega t'}&=&
\sum\limits_{n=-\infty}^\infty
J_n(-a){\e^{\ii n\omega t}-\e^{-ct}\over c+\ii n\omega}.
\end{eqnarray*}
Here, $J_n(a)$ are the Bessel functions of the first kind.
As we are interested only in the behaviour for large enough times (the steady state
solution),
we remove all terms that are damped out exponentially for $t\gg c^{-1}$ 
from the very beginning. Note,
that $c^{-1}$ is indeed the time scale for the transience, as has been claimed before.

For the mean velocity, we obtain
\begin{widetext}
\begin{align}
v(t)=h\cos\omega t+\eta^2\sum\limits_{m,n=-\infty}^\infty
\int{\cd q\over2\pi}(\ii q)\Delta(q)J_m\left(q{h\over\omega}\right)
J_n\left(-q{h\over\omega}\right)
{\e^{\ii(m+n)\omega t}(c-\ii n\omega)\over c^2+n^2\omega^2}
\end{align}
\end{widetext}
In principle, this is already a Fourier series representation, not very
elegant, though. The argument $(m+n)\omega t$ of the expansion basis 
exponentials promises a rather complicated structure for the coefficients.
A first observation, however, can already be made: Under the $q$ integral we
find an odd function $(\ii q)\Delta(q)$ and a product of two Bessel functions
of order $m$ and $n$, respectively. For the $q$-integral to result in a finite value, 
a function is required that is not odd in $q$. This necessitates the product
of the two Bessel functions to be odd, or, equivalently, $m+n$ to be an odd number.
Whence, we conclude, that to first perturbative order, our symmetry argument
(Fourier coefficients for even $N$ must vanish) is fulfilled exactly.

It requires some
tedious algebra to collect all contributions belonging to a
certain harmonic order from the double series. Eventually, we obtain
the series expansion
\begin{widetext}
\begin{align}
\label{eq:mf1sto}
{{v}(t)\over h}&=\cos\omega t+
\sum\limits_{N=1}^\infty\left[A_N\cos N\omega t+
B_N\sin N\omega t\right]\\
\label{eq:1stoan}
A_N&=2{\eta^2\over h^2}{\omega^2\over c^2}\left[
\sum\limits_{n=1}^{N-1}{(-1)^nnK_{N-n,n}\over 1+n^2\omega^2/c^2}
-\sum\limits_{n=1}^\infty{nK_{N+n,n}\over 1+n^2\omega^2/c^2}
-\sum\limits_{n=N}^\infty{nK_{n-N,n}\over 1+n^2\omega^2/c^2}
\right]+{\cal O}(\eta^4)\\
\label{eq:1stobn}
B_N&=2{\eta^2\over h^2}{\omega\over c}\left[
-\sum\limits_{n=1}^{N-1}{(-1)^n K_{N-n,n}\over 1+n^2\omega^2/c^2}
-\sum\limits_{n=0}^\infty{K_{N+n,n}\over 1+n^2\omega^2/c^2}
+\sum\limits_{n=N}^\infty{K_{n-N,n}\over 1+n^2\omega^2/c^2}
\right]+{\cal O}(\eta^4)
\end{align}
For convenience, we introduced the following abbreviation (depending only on
the parameter ratio $h/(\omega\ell)$), in which ${}_3F_3$
denotes the generalised hypergeometric function \cite{Gradshteyn}.
\begin{align}
\label{eq:mfkmnsym}
K_{m,n}=&{h\over\ell}\int{\cd q\over2\pi}\>q\Delta(q)\cdot 
J_m\left(q{h\over\omega}\right)J_n\left(q{h\over\omega}\right)\\
=&2\left({h\over\omega\ell}\right)^{m+n+1}{\Gamma([m+n+2]/2)\over
\sqrt\pi\cdot m!\cdot n!}\times\\
&{}_3F_3\left[\left\{{m+n+1\over 2},{m+n+2\over 2},{m+n+2\over 2}\right\},
\{m+1,n+1,m+n+1\},-{4h^2\over\omega^2\ell^2}\right].\nonumber
\end{align}
\end{widetext}
Note, that taking $\omega\to 0$ is forbidden here, as we used $\omega\ne 0$ while
deriving the coefficients and moreover perturbation theory breaks down (recall that
$h/\eta\gg c/\omega$). The same holds for $\ell\to 0$.
The remaining extreme limits $\omega\to\infty$ and 
$\ell\to \infty$ are not interesting, since in these limits the disorder is rendered 
unimportant. Therefore, in the following, we assume finite (positive) values for
$\ell$ and $\omega$ and moreover set them equal to one $\omega=\ell=1$,
by appropriately choosing the units for $z$ and $t$.


Now, we are left with three dimensionless parameters: 
$h$, $c$ and $\eta$.
The dependence of the first order perturbative Fourier coefficients on $\eta$ is trivial. 
The dependence on $c$
is also evident, as can be read off from (\ref{eq:1stoan},\ref{eq:1stobn}). For larger $c$,
the system is more tightly bound to the non-disordered solution, 
supressing perturbative corrections.

\begin{figure}
\includegraphics[width=\columnwidth]{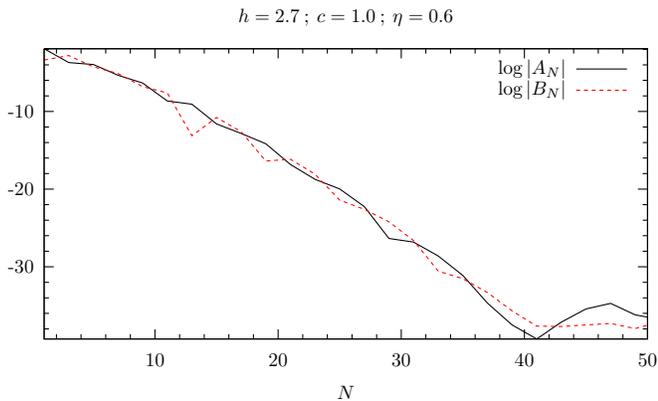}
\caption{\label{fig:habh1}%
Plotting the logarithms of $|A_N|$ and $|B_N|$ reveals
the exponential decay with $N$. In the regime where numerical errors do not
dominate the result, a linear regression seems appropriate.
}%
\end{figure}%
\begin{figure}
\includegraphics[width=\columnwidth]{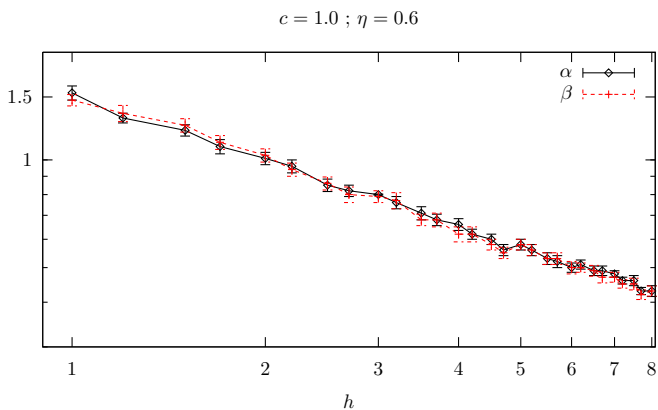}
\caption{\label{fig:habh2}%
Performing the
linear regression for many $h$ yields slopes $\alpha$ and $\beta$ appearing to depend 
on $h$ in a power-law fashion.}
\end{figure}%
The most interesting but also the most difficult is the dependence of the Fourier coefficients
on $h$. Actually, there are two competing effects. On the one hand, large driving strengths
render the disorder unimportant in all cases accessible through perturbative methods.
In a nutshell, the expansion parameter is $\eta/h$. On the other hand, 
if one thinks of $g(Z(t))$ as a function of time, the more rapid $Z(t)$ 
changes the more $g$ fluctuates on short time scales and thus
brings higher frequency contributions to $v(t)$. The first remark is reflected in
the overall weight of the Fourier coefficients as corrections to the
non-disordered case, decreasing with $h$. The second idea is expected to express itself 
in the decay
of the Fourier coefficients with $N$. The larger $h$, the weaker we expect this 
decay to be.

In equations (\ref{eq:mf1sto}) and (\ref{eq:mfkmnsym}), the 
dependence of the higher harmonics on $h$ is hidden in the $K_{m,n}$ as
functions of $h$:
The maximum of the $K_{m,n}$ as functions of the parameter $h/(\omega\ell)$
shifts to larger values as $m$ or $n$ increase. In the integral representation this
can be seen, as the Bessel functions take their first extrema at large arguments
for large indices. However, the complicated way in which the $K_{m,n}$ functions enter
$A_N$ and $B_N$ hinders an analytic access to the decay law.
A numerical determination of the Fourier coefficients for the perturbative result
reveals an exponential decay, cf. fig. \ref{fig:habh1}. The noisy behaviour for 
$N\ge 40$ is due to numerical fluctuations. Note, that these fluctuations are of
the order $10^{-14}$, which is quite reasonable. The plot in fig. \ref{fig:habh1}
is mere illustration of a more general phenomenon. This exponential decay
has been found for many sets of parameters, thus one is led to the ansatz
\begin{equation}
|A_N|\sim\e^{-\alpha N}\quad;\quad |B_N|\sim\e^{-\beta N},
\end{equation}
where $\alpha$ and $\beta$ can be estimated through a linear regression up to a suitable 
$N_{\text{max}}$. Of course, it is not expected, that $\alpha$ and $\beta$ are distinct,
nor that they depend on the parameters in different ways. Determining both just doubles
the amount of available data.

As our results are first-order perturbative, $\alpha$ and $\beta$ must not 
depend on $\eta$. The main interest now focusses
on the dependence of the decay constants on $h$. The results from a linear regression for
a series of $h$-values, $c$ and $\eta$ kept fixed, suggest a power-law dependence
\begin{equation}
\label{eq:regression}
\alpha(h)=C_\alpha\cdot h^{-\xi_\alpha},\quad\beta(h)=C_\beta\cdot h^{-\xi_\beta}.
\end{equation}
Fig. \ref{fig:habh2} displays this relation for a particular example. Repeating
this data collection and subsequent regression for different values for $c$ and $\eta$
yields the results summarised in table \ref{tab:habh}. 
\begin{table}
\caption{\label{tab:habh}Results for the regression (\ref{eq:regression}).}
\begin{ruledtabular}
\begin{tabular}{cccccc}
$c$&$\eta$&$C_\alpha$&$C_\beta$&$\xi_\alpha$&$\xi_\beta$\\
\hline
1.0&0.6&1.52&1.52&0.61&0.61\\
1.5&0.6&1.56&1.56&0.58&0.59\\
2.0&0.6&1.56&1.59&0.58&0.60\\
2.5&0.6&1.63&1.62&0.61&0.61\\
3.0&1.0&1.66&1.66&0.63&0.63\\
3.5&1.0&1.71&1.68&0.62&0.62\\
4.0&1.0&1.69&1.65&0.61&0.60\\
4.5&1.0&1.72&1.68&0.62&0.62\\
5.0&2.0&1.71&1.67&0.61&0.60\\
5.5&2.0&1.73&1.73&0.61&0.62\\
6.0&2.0&1.77&1.72&0.62&0.62\\
6.5&3.0&1.76&1.77&0.62&0.62
\end{tabular}
\end{ruledtabular}
\end{table}
While the exponent $\xi$ appears constant $\xi\approx 0.6$, the prefactor
seems to depend on $c$. 
\begin{figure}
\includegraphics[width=\columnwidth]{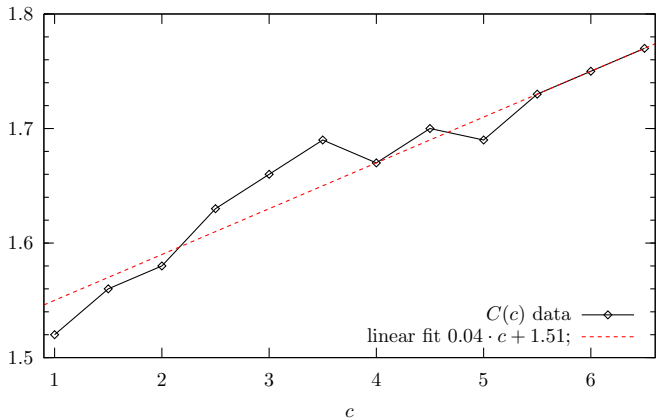}
\caption{\label{fig:vorfc}Plot of the change of the prefactor $C(c)$ in (\ref{eq:regression})
on $c$. The linear fit yields a fairly tiny slope.}
\end{figure}%
An attempt to redo the same procedure, done for $h$,
with the parameter $c$ to gain information
about the functional dependence of $\alpha$ and $\beta$ on $c$ yields a complicated
but rather weak dependence, which gives no further insight. The obvious approach is
to visualise the dependence of $C_\alpha$ and $C_\beta$ on $c$. The linear fit
in fig. \ref{fig:vorfc} gives a fairly tiny slope, so the dependence of the decay
constants on $c$ may be assumed to be weak.

%
Certainly, it is desirable to ascertain
the validity of this decay law beyond perturbation theory. In a few words, it ought to
be explained, why we have not been able to do it. First of all, the logarithmic plots
of the Fourier coefficients in fig. \ref{fig:habh1} exhibit fluctuations around the
linear decrease. This ``noise'' is authentic and not attributed to numerical
inaccuracies. The exponential decay of the Fourier coefficients is superimposed on
a true, complicated dependence. Hence, it requires a lot of data points to obtain
reasonable data. Since the Fourier coefficients for even $N$ vanish, 
in the example of fig. \ref{fig:habh1} the
regression can be carried out over around 15-20 data points. This is a fair number. The
quality relies heavily on the accuracy of the numerical determination of the
Fourier coefficients. In Fourier analyses of the numerics for the full equation
of motion (\ref{eq:acmfeom}), we did not manage to get a precision better than
of the order of $10^{-3}$. This means, the regression has to be stopped at 
$N_{\text{max}}$, where $\log A_{N_{\text{max}}}\approx -7$. In the example
of fig. \ref{fig:habh1}, this leaves us with less than 5 data points. In view of
the natural fluctuations, a linear regression is not sensible any more.

\section{\label{sec:conclusions}%
Conclusions%
}
The mean-field version of the problem of ac-driven elastic
interfaces in disordered media admits a
regular perturbative treatment that, where applicable, agrees very well
with the numerics for the full equation of motion. It has been shown,
that diagrammatic contributions with an unbounded increase in time
cancel among each other, leaving a well-behaved perturbative
expansion.

The solutions to the mean-field equation of motion are found to share
many features with the numerical solutions of the original problem,
like the hysteretic behaviour of the $v$-$h$-plots.

Unfortunately, the perturbative expressions are very complicated
and thus only of little use for analytic insights. However, they
improve numerical results tremendously which made possible to
establish the dependence of the decay constants of the Fourier
modes on $h$ as a power law $\alpha,\beta(h)=C(c)\cdot h^{-\xi}$
with $\xi\simeq 0.6$.

\begin{acknowledgments}
For fruitful discussions, inspiration and ideas I am grateful
to T. Nattermann. Further, I want to thank A. Glatz and
Z. Ristivojevic for discussions.
Finally, I would like to acknowledge financial support by 
Sonderforschungsbereich 608.
\end{acknowledgments}

\appendix

\section{\label{app:reg}%
Regularity of the perturbation expansion%
}
In section \ref{sec:pert:consistency} we have analysed, how the unbounded contributions,
contained in the two diagrams that involve a curly line, mutually cancel in the
second non-vanishing perturbative order. In this appendix, 
we are going to explain how this cancellation process generalises to all orders in 
perturbation theory. As before, for simplicity, 
we work with the diagrams for the disorder-averaged
velocity, that arise by
just removing the curly lines from the root of the diagrams for $\diszeta$
(cf. equation (\ref{eq:diagexpand})). 
In a velocity diagram contributing to the $n$-th order (recall, that only for even $n$
the corrections are non-zero), 
any curly line connects two trees of order $p$ and $q$ (both even) 
with the restriction $p+q=n$.
Both trees appear in the expansion of lower orders, 
namely $p$ and $q$, respectively.
In the following, we want to sketch an inductive proof for the claim
that the unbounded terms
originating from trees with curly internal lines cancel among each other.

Let us assume, that for order $n$ we have achieved to ensure regularity. 
For every unbounded tree $T$, there is thus a set $T^1,\ldots,T^a$ of, let us call them
\emph{cancelling trees}, such that $T+T^1+\ldots+T^a$ is a regular, 
bounded expression in time.
As a starting point for the induction, take $n=4$, where the validity of
the claim has been verified in section \ref{sec:pert:consistency}.
It is now the task to validate the regularity for order $n+2$. 
First of all, we consider the process of attaching the root of a
regular tree $S$ (with no internal curly line) of order $s$ 
by a curly line to a vertex $v$ of another regular tree $R$ of order $r=n+2-s$ 
to obtain a new irregular tree $A$ of order $n+2$.
The vertex $v$ must be connected to another vertex $w\in R$ by a dashed line, to
carry out the Gau\ss ian disorder average. Without loss of generality, we assume
that $v$ is connected to $w$ by a path that first makes a step towards the root.  
The rules for the diagrammatic
expansion ensure, that there is a maximal regular subtree $T\subset R$, 
which contains $v$ and $w$.

Using partial integration, it is possible to move the vertex to which $S$ is connected
(via the curly line) to a neighbouring vertex in $T$. Thus, it is possible to
move the connection vertex along the unique way (in $T$) from $v$ to $w$. 
We are going to show, that once $w$ is reached, we have
obtained the cancelling tree which is unique. 
Diagrammatcially, the
process of moving the connection vertex from $v$ to $w$ reads:
\begin{eqnarray*}
\parbox{21mm}{%
\includegraphics{cdiag.1}%
}
&=&
\>\>\parbox{21mm}{%
\includegraphics{cdiag.2}%
}+D
\end{eqnarray*}
Here, the blank circle represents $S$, the lightgrey circle stands for the subtree $R_1$ of 
$R$, to which $v$ connects and the 
darkgrey shaded circle denotes trees which run out of $v$ 
(summarised in the following as $R_2$). Certainly, in general there 
may be dashed lines between the dark- and the lightgrey circle, which we have omitted as
they are not relevant for the forthcoming discussion.
The dotted line just serves as a joker - it is not important to specify how many trees
go out of $v$. 
The last term $D$ collects the left-over terms
from the partial integration. Note, that, if it takes several steps to go from $v$ to $w$,
the intermediate expressions (in the partial integration) are no valid diagrams.

To illustrate the procedure, we take a look at the first step:
\begin{widetext}
\begin{eqnarray*}
\parbox{21mm}{%
\includegraphics{cdiag.1}%
}
&=&R_1(t)\int\limits_0^{T_1}\cd t_1\>\e^{-c(T_1-t_1)}(-1)^\nu\Delta^{(\mu+\nu)}[Z(\tau)-Z(t_1)]
R_2(t_1)\int\limits_0^{t_1}\cd t_2 S(t_2)\\
&=&R_1(t)\int\limits_0^{T_1}\cd t_2 S(t_2)
\int\limits_0^{T_1}\cd t_1\e^{-c(T_1-t_1)}(-1)^\nu\Delta^{(\mu+\nu)}[Z(\tau)-Z(t_1)]R_2(t_1)\\
&&-R_1(t)\int\limits_0^{T_1}\cd t_1\>\e^{-c(T_1-t_1)}S(t_1)\int\limits_0^{t_1}
\cd t_2\>\e^{-c(t_1-t_2)}(-1)^\nu\Delta^{(\mu+\nu)}[Z(\tau)-Z(t_2)]R_2(t_2)
\end{eqnarray*}
\end{widetext}
The order of the derivative (i.e. the number of outgoing lines) of $w$ and $v$ are denoted 
by $\mu$ and $\nu$, respectively.
The time, at which the whole diagram is to be evaluated, is $t$, the time corresponding to
the vertex to which $v$ is connected is given by $T_1$, $t_1$ is thus the time associated 
to $v$ and so on. The time of $w$ is $\tau$. 
Thus, we see, that if $w$ is not the vertex to which $v$ is directly connected 
(then $T_1\ne\tau$ in general),
the first expression after partial integration cannot be a valid diagram: $v$ has lost
one order of derivative ($\nu-1$ lines go out instead of $\nu$), but the derivative
of the correlator $\Delta$ has not changed. A valid diagram is then obtained, when the
connection of $S$ has reached $w$. Then, $v$ has lost an outgoing line, but $w$ received
one more and we indeed have achieved a cancelling tree:
the factor $(-1)^\nu$ remains, the true diagram, however, has $(-1)^{\nu-1}$. The signs are
different, thus the two trees cancel. 
The left-over term from the partial integration is again regular, as can be seen since
all time integrals carry an exponential damping term. It is clear, that this is generally 
true for every partial integration step.

To go one step further, we assume now $S$ to be irregular.
Essentially, the same procedure works, but there are more
cancelling trees: one has take all cancelling trees $\{S^i\}$ for $S$ into account 
(which exist by induction hypothesis), thus
$S$ is replaced by $\sum S^i$ and thence the left-over terms are again regular.

A possible irregularity of $R$ can be accounted for in the same way. It is, however,
important to explain why this is possible, i.e. what is $v$ and $w$ in the cancelling
trees for $R$. In the case of irregular $S$ the problem was easy, since all
trees have a unique root. As we have seen already, the procedure of creating cancelling
trees does not change the structure of regular subtrees. Thence, all cancelling trees for 
$R$ contain $T$. This makes clear, which $v$ and $w$ have to be chosen in the cancelling
trees: they are well-defined in $T$ and $T$ is a well-defined subtree of the cancelling
trees. Thus, repeating the whole procedure described above for all cancelling trees of $R$
yields the complete set of cancelling trees for $A$ in the most general setting.

\section{\label{app:resum}%
Resummation of the perturbation expansion%
}
For a constant driving force, it is not clear, why a perturbative
approach using (\ref{eq:mfdisexpansion}) should work.
So,
we approach the perturbation expansion from another direction, which
will turn out to be a resummation of our former expansion in the 
disorder.
Take our original equation of motion
\begin{align}
\partial_tz&=c\cdot\left[\disav{z}-z\right]+h\cdot f(t)+
\eta\cdot g(z)
\end{align}
and decompose $z=X+\xi$, where $X=\disav{z}$. Thus $\disav{\xi}=0$.
This gives us two non-linearly coupled non-linear
differential equations
\begin{align}
\partial_tX(t)&=h\cdot f(t)+\eta\cdot\disav{g(X+\xi)}\\
\label{eq:xidgl}
(\partial_t+c)\xi(t)&=\eta\big[g(X+\xi)-\disav{g(X+\xi)}\big]
\end{align}
If $c$ is large enough, one can always achieve $\xi\ll\ell$ and
thus a Taylor expansion of the disorder around $X(t)$ keeping
only lowest order-terms seems reasonable.
Instead of a systematic expansion in the disorder, we now perform
power-counting in $\xi$. This leads to a recursive structure
for $\xi$:
\begin{align*}
(\partial_t&+c)\xi=\eta\sum\limits_{n=0}^\infty{1\over n!}\left[
g^{(n)}(X)\xi^n-\disav{g^{(n)}(X)\xi^n}\right]\\
&=\eta\left[g(X)+g'(X)\xi-\disav{g'(X)\xi}+{1\over 2}g''(X)\xi^2+\ldots\right]
\end{align*}
and thus a self-consistent equation for $X(t)$
\begin{widetext}
\begin{align*}
\partial_tX&=h\cdot f(t)+\eta\sum\limits_{n=0}^\infty{1\over n!}\disav{g^{(n)}(X)\xi^n}\\
&=h\cdot f(t)+\eta
\Big\langle g(X)+\eta g'(X)\int\limits_0^t\cd t'\e^{-c(t-t')}
\big[g(X)+\ldots\big]+
\ldots\Big\rangle
\end{align*}
The graphical structure is now similar to that of section \ref{sec:pert:diag}.
It reads
\begin{align}
\label{eq:diagXexp}
\partial_tX=&h\cdot f(t)+\>
\parbox{9mm}{\includegraphics{mvdiag.1}}\>+3\cdot\>
\parbox{9mm}{\includegraphics{mvdiag.2}}\>+\>
\parbox{17mm}{\includegraphics{mvdiag.3}}\>+\>
2\cdot\>\parbox{17mm}{\includegraphics{mvdiag.4}}\>+\nonumber\\
&\nonumber\\&
2\cdot\>
\parbox{17mm}{\includegraphics{mvdiag.5}}\>+2\cdot\>
\parbox{17mm}{\includegraphics{mvdiag.6}}\>+\>
\parbox{25mm}{\includegraphics{mvdiag.7}}\>+\>
\parbox{25mm}{\includegraphics{mvdiag.8}}\>+
{\cal{O}}(\xi^4)
\end{align}
\end{widetext}
This series only consists of ``irreducible'' (1PI) graphs, where no line can
be cut such that the whole graph falls apart into two and 
no curly lines occur. Otherwise the diagrammatic rules are essentially 
the same as before. Every vertex corresponds to $g^{(m)}(X)/m!$, where $m$ counts 
the number of outgoing lines. The full lines are propagators of the
differential equation (\ref{eq:xidgl}) for $\xi$.

On analysing the expansion of the first and simplest diagram in the disorder
(up to ${\cal O}(\eta^6)$)
\begin{align*}
\parbox{9mm}{\includegraphics{demvdiag.1}}\>
&=\>
\parbox{9mm}{\includegraphics{demvdiag.2}}\>+2\cdot\>
\parbox{17mm}{\includegraphics{demvdiag.3}}\>+\>
\parbox{25mm}{\includegraphics{demvdiag.4}}\\
g(X)&=g(Z+\diszeta)=g(Z)+g'(Z)\diszeta+\ldots,
\end{align*}
one inspects that the ``new'' diagrammatic expansion (\ref{eq:diagXexp}) is merely
a resummation of our old one. However, if a constant driving force is exerted,
$\xi(t)$ remains bounded (and, depending on $c$, also small) 
at all times, in contrast to $\zeta(t)$. This can be seen by
confirming the earlier estimate for the mean deviation of a realisation
from the mean position (cf. section \ref{sec:model}), which is exactly $\xi$. 
We observe that
\begin{align}
\disav{(\disav{z}-z)^2}&=\disav{\xi^2}
={\eta^2\over c^2}\cdot\phi\left({c\ell\over 2v}\right)+{\cal O}(\eta^4).
\end{align}
As $\phi\le 1$, to this order $\eta/c$ is an upper bound for the
typical distance from the mean. Thus this
perturbation expansion should work fine for $\eta/c\ll\ell$.
As far as there is overlap, this also agrees with the result in \cite{KL:PRB85}.
For a constant driving force $h$, the first order correction to the velocity yields
a self-consistent integral equation
\begin{equation}
\label{eq:dcv1stoscp}
v=h+\eta^2\int\limits_0^t\cd t'\>
\e^{-c(t-t')}\Delta'[v(t-t')]+{\cal O}(\eta^4),
\end{equation}
which has been used in computing equation (\ref{eq:pert4o}).

\bibliography{wand}

\end{document}